# Photonic integrated circuit with multiple waveguide layers for broadband high-efficient on-chip 3-D optical phased arrays in light detection and ranging applications

DACHUAN WU[1], BOWEN YU[1], VENUS KAKDARVISHI[1], AND YASHA YI[1,2,*]

[1]Integrated Nano Optoelectronics Laboratory, University of Michigan, Dearborn, MI, 48124, USA
[2]Energy Institute, University of Michigan, Ann Arbor, MI, 48109, USA
*yashayi@umich.edu

**Traditional photonic integrated circuit (PIC) inherits the mature CMOS fabrication process from the electronic integrated circuit (IC) industry. However, this process also limits the PIC structure to a single-waveguide-layer configuration. In this work, we explore the possibility of the multi-waveguide-layer PIC by proposing and demonstrating a true *3-D optical phased array* (OPA) device, with the light exiting from the edge of the device, based on a multi-layer $Si_3N_4/SiO_2$ platform. The multi-waveguide-layer configuration offers the possibility of utilizing edge couplers at both the input and the emitting ends to achieve broadband high efficiency. This uniqueness provides the potential for a more extended detection range in the Lidar application. The device has been studied by numerical simulation, and proof-of-concept samples have been fabricated and tested with a CMOS-compatible process. To the best of our knowledge, this is the first experimental proof-of-concept of a true *3-D* OPA with a multi-waveguide-layer configuration all over the device.**

## INTRODUCTION

During the development of the electronic integrated circuit (IC) industry, photonic integrated circuits (PIC) have been proposed as the next-generation chips and studied for decades. Normal PICs inherit the mature CMOS fabrication process from electronic IC; they usually have a single waveguide layer on the top of a silicon-on-insulator (SOI) platform or are based on deposited silicon nitride ($Si_3N_4$). Usually, the fabrication uses the top layer as the waveguide layer, and then the electronic contacts are fabricated above the waveguides for the modulation. While this technique takes some advantages from the mature CMOS fabrication process, it restricts the PICs in the single-waveguide-layer configuration, limiting the device performance. In recent years, the electronic IC industries have shown a trend of converting their memory and computing unit designs from 2-D to 3-D [1-2]. Nevertheless, these fabrication processes can also be applied to 3-D multi-waveguide-layer PICs. A relatively new type of PIC, called optical phased array (OPA), has drawn much research attention due to its potential in LiDAR applications. Yet, this device suffers from the limitation of single-waveguide-layer configuration.

*Light Detection and Ranging* (LiDAR) systems are used primarily for full dimensional sensing, with applications ranging from navigation for autonomous vehicles to robotics, imaging, unmanned aerial vehicles (UAVs), national security, healthcare, and the Internet of Things (IOTs) [3-4]. With the time of flight (ToF) or frequency modulated continuous wave (FMCW) mechanism, a LiDAR system can generate a 3-D map of its surrounding with distance and velocity information [5-6]. Compared to the common mechanical LiDAR, which is usually high cost and slow in scanning [7-8], a chip-scale LiDAR system can provide both increased range and resolution required for high-speed driving—and other tasks, such as real-time facial recognition—that are beyond the capability of current LiDAR systems. With the growing interest from the research community in chip-scale LiDAR, beam steering based on the *integrated optical phased array* (OPA) has drawn a lot of research effort in the past decade [9].

Significant progress has been achieved, including thermal tuning [10-11], electro-optics tuning [12], high sensitivity wavelength tuning [13-14], integrated on-chip light source [8, 15], and side lobe suppression by aperiodic or apodized array placement [21-23]. However, most on-chip OPA research stays in the single-waveguide-layer structure [9-21]. The OPA formed by a single layer can only emit the beam by diffractive components such as grating couplers, which has narrowband and relatively low emitting efficiency. In previous work [24], we showed that about half of the light could be emitted to the substrate in a normal grating-coupler-based OPA. The optical efficiency of the beam steering devices is directly related to the detection range of LiDAR, and most applications (particularly for AD and ADAS) require a detection range to be at least 100 meters. Designs can be applied to suppress substrate leakage of the energy [25, 26], but only benefits in a narrow bandwidth.

Previous works have attempted to address the relatively *low emitting-efficiency* challenge [27–32]. The basic idea is to use end-fire emitters to achieve high efficiency [27, 28]. Further works aiming to confine the waveguide spacing to half-wavelength have been done using various approaches [29, 30]. These works employ the configuration with a single waveguide layer and thus offer the convenience of tuning the phase of each waveguide [27, 29, 30]. Unfortunately, the beam emitted by such a configuration is a fan-beam with only 1-D convergence, as the single waveguide layer can only form a 1-D OPA on the edge of the chip. The possibility of emitting a 2-D converged beam from the edge (end-fire) requires a true 3-D OPA on the edge side. This is discussed in [30] and [31]. In [31], the idea of a true 3-D OPA is firstly proposed; the performance of an end-fire OPA with a multi-waveguide-layer configuration is numerically discussed, and the method utilizing nanomembrane transfer printing to fabricate a multi-layer structure with the top Si layer from an SOI wafer is proposed. In [32], a direct writing method based on ultrafast laser inscription (ULI) is applied to achieve the conversion between single-layer waveguides and 3-D waveguides in the structure; therefore, we know that a 3-D OPA can be formed on the edge side, especially with the current *CMOS compatible 3D circuitry on-chip* [33]. This is the first reason that a multi-waveguide-layer configuration is helpful in an OPA device.

Another major issue for the optical efficiency occurs at the input coupling end. In most OPA studies, researchers considered using an external light source such as a pulsed or coherent laser. In such a case, fiber is needed to connect the light source and the OPA chip. Many research shows that the frequency of light is usually utilized as one degree of freedom in either beam steering or distance detection (in the case of an FMCW Lidar). Therefore, a wideband coupling performance is desired for the fiber-to-chip coupler. Similarly, at the emitting end, the two standard coupler designs also have their issues: the edge coupler shows a wideband performance but usually suffers a significant coupling loss due to the mode mismatch between single-mode fiber and on-chip waveguide (which is generally at least one order smaller in size than the fiber) [34]; the grating coupler can offer a much better mode match, which leads to a high coupling efficiency, but only at a relatively narrower band [35]. To address this issue, wideband high-efficient couplers have been designed by applying additional waveguide layers (usually $Si_3N_4$ layers) on top of the silicon waveguide layer. They allow the fiber mode to be first coupled to a super-mode in the nitride stack, then gradually coupled to the silicon waveguide by evanescent coupling [36, 37]. This is the second reason multi-waveguide-layer configuration has its advantage in an OPA device.

The edge coupler can offer a stable efficiency in a broad bandwidth. However, in the particular case of OPA devices, trivially applying the edge coupler will result in two disadvantages: 1. a mode mismatch at the input end, and 2. a beam with only 1D convergence at the emitting end. In this work, we have made one step forward to address both disadvantages. A *true 3D Optical Phased Array* has been demonstrated for beam steering by applying the multi-waveguide-layer configuration to the whole device. It takes advantage of building PIC in 3-D [33], based on *a novel $Si_3N_4/SiO_2$* platform, and achieves distinct characteristics: *a)* Vertical multiple-layers provide a broadband high power coupling efficiency to a 2-D converged beam.

*b)* The Ω-shape design purposely creates an extra dispersion effect, which enables an unlimited beam steering capability in principle. As a result, the highest proven beam steering is 0.577°/1nm wavelength in the 4-layer sample with $60 \mu m$ delay length. This differs from all other standard 2D approaches leveraging grating couplers to synthesize 3D beams. As a proof of concept, we experimentally fabricated the multilayer OPAs and verified the idea with careful characterization.

## RESULTS
### Device Structure

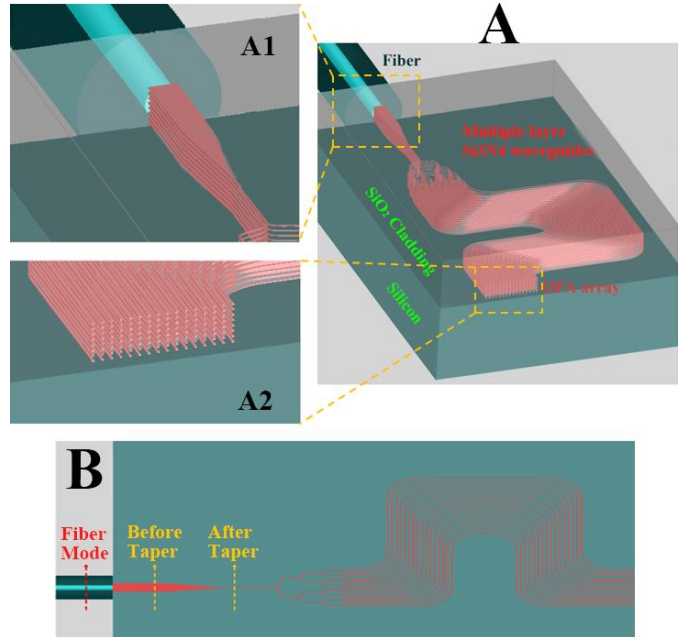

**Figure 1. Illustration of the multi-layer Si$_3$N$_4$ 3D OPA.** (**A**) Schematic (3D view) of the structure. An SMF is used to couple light into the device, and the waveguide width at the coupling region is enlarged to ensure the best mode matching (see zoom-in figure **A1**); a 2D 8X16 OPA is formed at the edge of the device (see zoom-in figure **A2**). (**B**) Top view layout of the waveguide layers. The mode matching at the input coupler region is illustrated in Fig. 2.

The proposed device is illustrated in Fig. 1. A single-mode fiber (SMF-28) and an on-chip edge coupler are utilized to couple the light from the source (a tunable laser) to the waveguides. The mode match is supported by multiple waveguide layers to maximize the coupling efficiency. In addition, the thickness of Si$_3$N$_4$ waveguide layers and SiO$_2$ isolation layers are selected to minimize vertical crosstalk. A taper waveguide is applied to convert the mode size into single-mode waveguides. The Y-splitter tree is then used to split the light into multiple channels (16 channels for every layer in Fig. 1), and the phase in each channel is the same. The Ω-shape delay line region is designed to enable the beam steering capability. To achieve such functionality, a certain delay length is applied between each waveguide, introducing an extra artificial dispersion effect so that the emitting beam can be steered horizontally by wavelength tuning. At the emitting end, a 2D end-fire array is formed on the side of the device; the light is edge-coupled into the free space. The emitting efficiency is relatively high in a broad bandwidth due to the edge coupling. In short, the 2D array shapes the light beam in the following manner: in the horizontal direction, the Ω-shape delay line region controls the phase profile of the array; and in the vertical direction, the phase profile inherits the profile from the input fiber. We purposely design every layer to be in the same pattern, so there's no phase difference between layers. Si$_3$N$_4$ and SiO$_2$ are selected to be the waveguide and cladding material considering both device performance [39, 40] and fabrication possibility.

As stated above, there are two disadvantages when applying the edge coupler to a single-layer OPA in a trivial way: the mode mismatch at the input end, and the non-convergence in the vertical direction at the output end. They are addressed in this work by utilizing the multi-waveguide-layer configuration over the whole device. In the following sections, we firstly introduce the design of the input coupler, with the mode match supported by the multi-waveguide layer, then present the experimental results, which is a proof-of-concept to show the broadband high efficiency and the 2D converged beam.

### Design of the Input Coupler

The input fiber used in this work is SMF-28-J9 (Thorlabs), which has a mode field diameter (MFD) of $10.4\mu m$. The layer thickness of the device is optimized to minimize the vertical crosstalk. With the selected layer thickness, it can be calculated that eight waveguide layers can cover the full MFD. We firstly analyze the mode profile for three positions: the fiber, the coupler before taper, and the single mode waveguide after taper; these positions have been labeled in Fig. 1B.

As illustrated in Fig. 2, both TE and TM polarizations have been simulated. The comparison between B&E and C&F shows that the TM polarization has more field distribution into the cladding $SiO_2$ layers; this is better for the fiber to waveguide coupling as it can offer a better mode match. However, this mode profile is not wanted in two aspects: firstly, it will increase the optical loss for all the components, including waveguide, bending, and splitters; secondly, it does not help minimize the vertical crosstalk. On the other hand, the energy intensity of TE mode is more confined in the waveguide material, which is desired in this device. Therefore, TE polarization is selected in this work.

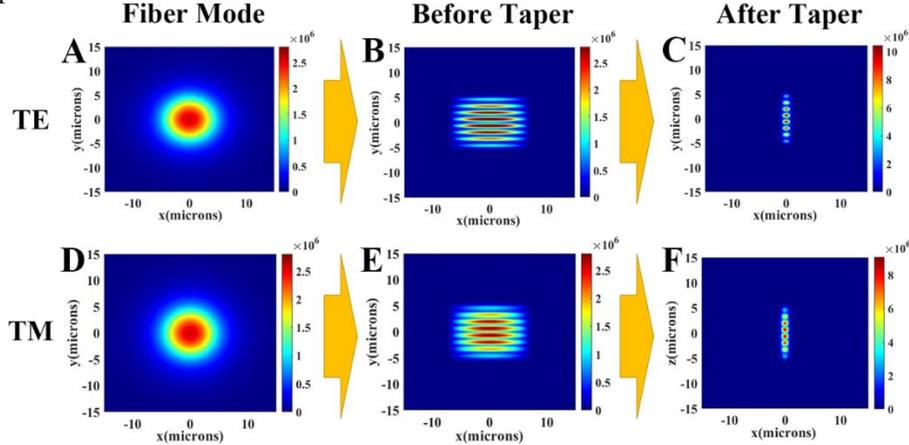

**Figure 2. Mode matching at the input coupler.** (**A**, **D**) Mode profile in the fiber for TE and TM polarization. (**B**, **E**) Mode profile before the taper for TE and TM polarization. (**C**, **F**) Mode profile after the taper for TE and TM polarization. TM polarization is more into the cladding layer, which is more suitable for input-coupling but results in a more considerable propagation loss and vertical crosstalk.

The width of the on-chip coupler (labeled as *before taper* in Fig. 1B) and the taper length is then optimized at the wavelength of 1550nm with TE polarization. We purposely designed every layer to be in the same pattern so that the phase profile across the OPA can inherit the profile from the fiber. Thus, the coupler width for every layer is the same in this design. Fig. 3A shows the sweeping results for the coupler width; the first coarse sweeping is done with a step of $0.5\mu m$, then a fine sweeping is followed with a step of $0.1\mu m$ at $13.5\mu m$ to $14.5\mu m$, the results show that a coupler width (labeled as *before taper* in Fig. 1B) of $13.9\mu m$ offers the best mode match, the coupling efficiency from fiber to coupler is 74.23%. Two factors contribute to the coupling loss: firstly, Fresnel reflection occurs on the interface between fiber and the device; and secondly, the layer thickness is selected to minimize the vertical crosstalk; this selection also makes that the mode profile is nearly zero at the center of every insulation $SiO_2$ layer, which results in partial mode mismatch. Fig. 3B shows the sweeping results of the taper length; the vertical axis in this figure is the total coupling efficiency of the edge coupler. The result curve gradually converges to 73.67% when the taper length approaches $400\mu m$; this corresponds to a 99.25% taper efficiency (73.67% / 74.23% = 99.25%). In this work, the taper length is selected to be $150\mu m$; the corresponding coupling efficiency is 71.24%.

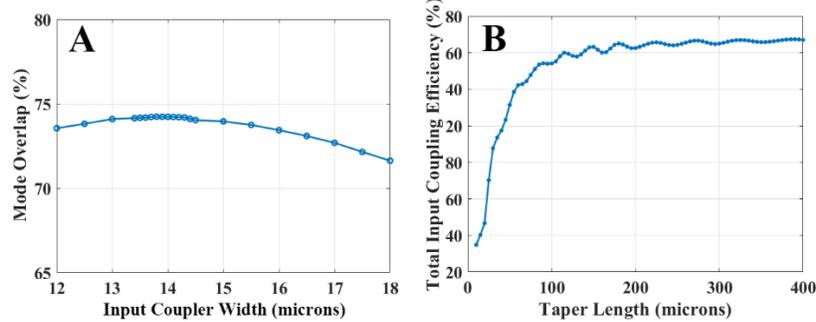

**Figure 3. Optimization of the input coupler.** (**A**) Coupler width (labeled as *before taper* in Fig. 1B) sweeping for mode matching between SMF and input coupler. With the pre-determined thickness of the waveguide layer and cladding layer, a waveguide width of $13.9\mu m$ offers the best coupling efficiency of 74.23% (this is the efficiency from fiber to input coupler). (**B**) Taper length sweeping for mode size conversion between input coupler (labeled as *before taper* in Fig. 1B) and single mode waveguide (labeled as *after taper* in Fig. 1B). The coupling efficiency converges to 73.67% at $400\mu m$ taper length (this is the efficiency from fiber to single-mode waveguide).

The mode propagation in the coupling region is plotted in Fig. 4, 4A shows the side-view cross-section; thanks to the optimized layer thickness, the light propagation in all the eight layers is individual; 4B shows the top-view cross-section, the mode size conversion between the fiber and the single mode waveguide can be observed. As stated above, the coupling efficiency from an edge coupler is relatively stable in a broad bandwidth; the coupler is optimized at the wavelength of 1550nm, and the coupling efficiency is tested over the wavelength region of 1500nm to 1600nm. Fig. 4C shows the results, the highest efficiency is 71.70% at 1515nm, and the efficiency variation in the whole range is only 0.58%.

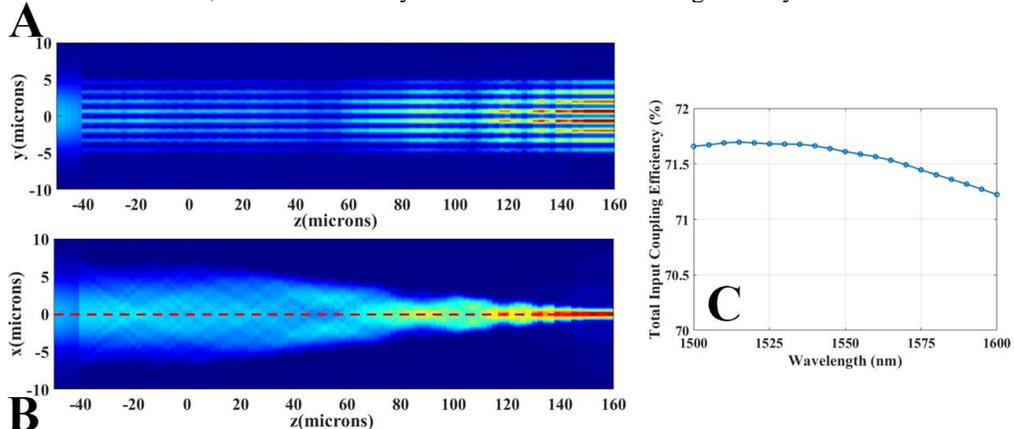

**Figure 4. Performance of the whole input coupler.** (**A**) Mode propagation across the entire input coupler from cross-section view (red dashed line in 4B) at 1550nm. The light propagates individually along every layer. (**B**) Mode propagation in the whole input coupler from the top view. (**C**) The simulated coupling efficiency of the entire input coupler at 1500nm to 1600nm wavelength range. The maximum efficiency appears at 1515nm to be 71.70%. Note this is the result without anti-reflection coating.

## Experimental Proof-of-Concept

The samples are fabricated in the Lurie Nanofabrication Facility (LNF) in Ann Arbor, Michigan, USA. Samples with 1 to 4 waveguide layers have been fabricated to prove the concept. Due to the fabrication capability, we can only fabricate samples with up to 4 waveguide layers for now. Despite the limited fabrication capability, the experimental results are still sufficient in comparing the single-layer and multi-layer configurations. In this part, we will present the experimental results, which clearly show that the 4-layer sample is better than the 1-layer sample in the coupling efficiency and beam convergence.

Fig. 5 shows the fabricated 4-layer sample; the dedicated backscattering SEM imaging takes the picture. From the image, part of the Y-splitter tree and the Ω-shape delay line region can be distinguished. The image is taken from an angle so that the end surface is presented, and the tooth-like shape at every pitch is the result of the PECVD cladding on a high aspect-ratio grating. The zoom-in picture shows one horizontal pitch; the four darker boxes are the emitting surface of the $Si_3N_4$ waveguides at different layers. Note that the SEM image is taken from an angle so that the $Si_3N_4$ waveguides become unclear from top to bottom.

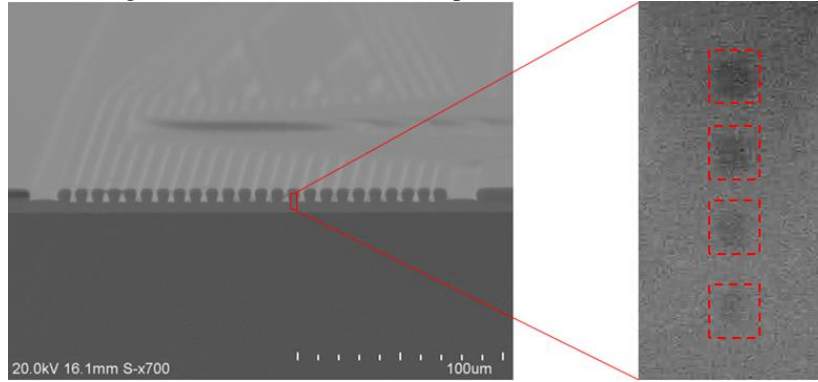

**Figure 5. SEM picture for a 4-layer sample. Left**: zoom-out view of the device from an angle. The Y-splitter tree and the delay line region can be distinguished from the picture. **Right**: zoom-in view of one pitch, four $Si_3N_4$ end-fire emitters can be distinguished, the image is taken from an angle, which results in the unclearness from top to bottom.

The input coupling efficiency is measured with the testing samples in the wavelength range from 1530nm to 1600nm (the source wavelength range is the C+L band, which is generated from Thorlabs TLX1 and TLX2). Fig. 6 plotted the tested efficiency. Compared to the typical Gaussian-curve spectrum from a grating coupler, the tested curves for the 1-layer to 4-layer devices are relatively flat; this is due to the edge coupler, which couples the light by direct mode match, but not the harmonic wave match (which is the case of the grating coupler). As a result, the input coupling efficiency for the 1-layer device is average -8.12 dB with a variance of 0.09 $dB^2$, and the efficiency is improved to an average of -4.57 dB with a variance of 0.13 $dB^2$ for the 4-layer device. Note these tested values includes the taper efficiency. These results show that the multi-waveguide-layer configuration can enhance the fiber-to-chip coupling efficiency. The fluctuation of the curves should be mainly due to the operation variations in the experiment. Compared to the simulated results (average of -7.58 dB for 1-layer structure and -2.64 dB for 4-layer structure), the tested efficiency of the 1-layer sample is 0.54 dB lower than the simulated value; this is because of the extra-waveguide loss due to the waveguide layer roughness from the fabrication error. This issue is more vital in the 4-layer device, as the roughness of the layers accumulates in the sample, which results in a more considerable propagation loss in the 4-layer device. Therefore, the tested input coupling efficiency of the 4-layer sample is 1.93 dB lower than the simulated value. It can be expected that this issue will be minimized with the state-of-the-art deposition method in an advanced foundry.

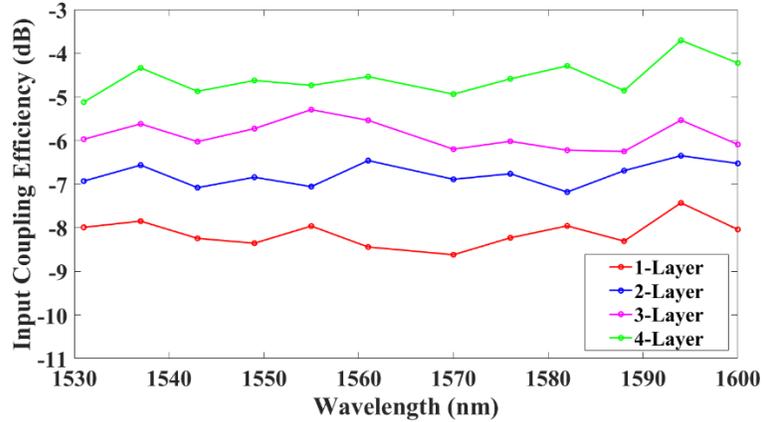

**Figure 6. Testing results of the input coupling efficiency.** Testing structure with straight waveguide and symmetric coupler is utilized to test the efficiency. The testing is done at the C+L wavelength band (1530nm to 1600nm) with laser sources TLX1 and TLX2 from Thorlabs.

The fiber-to-chip and emitting coupling are the two significant optical losses in a standard OPA device. As stated above, we utilize edge couplers over the whole device to achieve high efficiency at both ends. The multi-waveguide-layer configuration can enhance the mode match between the fiber and the coupler at the input end. Each waveguide layer matches a part of the fiber mode, so the total mode match is good. The light propagates individually at each layer, then emits through the end-fire emitter, interferes with the light from other layers in the free space, and eventually forms one or several (depending on the aliasing effect) beams. At the input end, the edge coupler relies on the multi-waveguide-layer configuration to achieve a good mode match. At the output end, the light is coupled from OPA to the free space, so there is no issue of mode mismatch. Compared to the grating couplers, which generate a considerable substrate leakage [24], using edge couplers will increase the emitting efficiency to approximately 70% (-1.55dB) [38], regardless of how many waveguide layers are in the sample. Note this efficiency is obtained when the device has no antireflection coating, so the Fresnel reflection produces that 30% loss. Fresnel reflection can be suppressed by anti-reflection coating; in principle, a correct anti-reflection coating can increase the theoretical emitting efficiency to nearly 100%.

The issue at the emitting end is the beam convergence. As proved in the previous studies [27, 29, 30], an end-fire OPA with a single-waveguide-layer configuration essentially means a fan-beam with only 1D convergence. In such a case, even though the OPA can have high optical efficiency, the beam's energy will be distributed in a vertical line with about 35° FWHM (according to the simulation), which means that the optical efficiency at one particular angle is still low. This issue is also addressed in this work by the multi-waveguide-layer configuration.

Fig. 7 shows the tested farfield pattern at 1550nm wavelength; samples without the Ω-shape delay line region are used for this testing, as they consistently emit the main lobe to the normal direction. 7A shows the farfield pattern for the 1-layer sample, and 7B shows the pattern for the 4-layer sample. It can be observed that the beams in 7B have apparent better vertical convergence than 7A. This is clear evidence to show the multi-waveguide-layer configuration address the vertical convergence issue. In the 1-layer sample, the emitting aperture is approximately 600nm. On the other hand, all waveguide layers have the same pattern in the 4-layer sample, which ensures no extra phase difference between layers. Consequently, the OPA emits the same vertical phase profile as the fiber mode, with only slight variation caused by the fabrication errors. Meanwhile, the emitting aperture has been increased to approximately $4.5\mu m$, which is 9 times greater than the 1-layer sample. 7C and 7D show the vertical cross-section of the samples, the red curve in 7C and blue curve in 7D are the simulated results, and the yellow curve in 7C and the green curve in 7D are the tested results. A good fit can be observed between the simulated and experimental results, which validate the effectiveness of the design. The tested FWHM for the 1-layer sample is 37.64° (simulated result is 32.12°), and for the 4-layer sample is 17.42° (simulated result is 14.26°). Based on the simulation result, a complete 8-layer device should be able to offer a vertical FWHM of approximately 5° [38]. It is also worth mentioning that the aliasing effect can be observed in Fig. 7A and 7B; those grating lobes appear at approximately ±11.20°, which agrees with the horizontal pitch of 8$\mu m$ that to be used in this work. In addition, it is possible to expect that an aperiodic design can be applied to suppress the grating lobes.

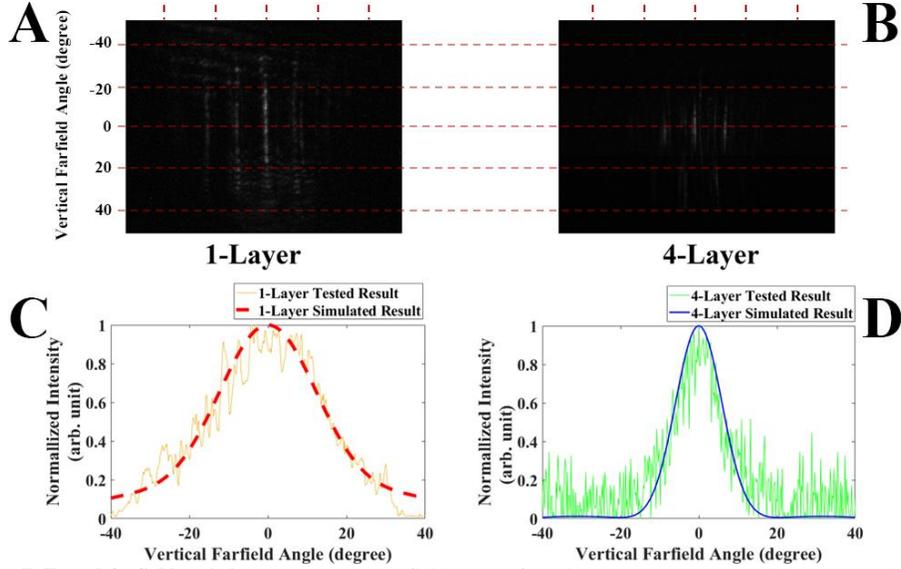

**Figure 7. Tested farfield emitting pattern.** (**A**) Farfield pattern for 1-layer structure, grating lobes can be observed due to the 8$\mu m$ horizontal pitch. (**B**) Farfield pattern for 4-layer structure, the vertical convergence of the main lobe and grating lobes are better than the 1-layer structure. (**C**) Vertical cross-section of the main lobe for the 1-layer structure. The simulated result shows an FWHM of 32.12°, and the tested FWHM is 37.64°. (**D**) Vertical cross-section of the main lobe for the 4-layer structure. The simulated result shows an FWHM of 14.26°, and the tested FWHM is 17.42°.

The beam shaping and steering in the horizontal direction are achieved by the Ω-shape delay line region. This design introduces an extra artificial dispersion effect into the device, which enables a highly sensitive beam steering capability. We have proven in our previous study [38] that the beam steering capability is linearly dependent on the length of the delay line, which the following equation can summarize.

$$\Delta\delta = \Delta\lambda * DL * b$$

where $\Delta\delta$ is the beam steering angle, $\Delta\lambda$ is the wavelength change, $DL$ is the length of the delay line, and $b$ is the essential steering sensitivity, which depends on the waveguide dimension.

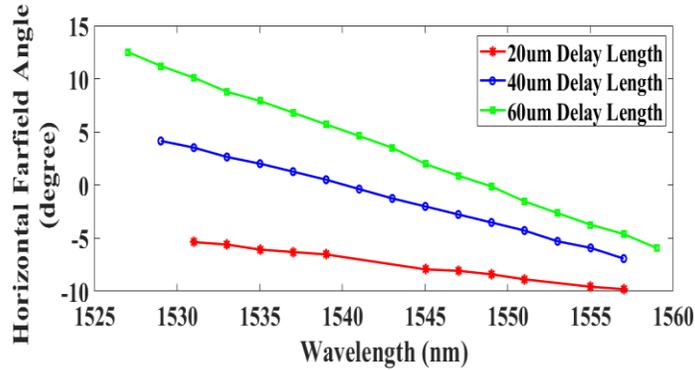

**Figure 8. Beam steering capability of the 4-layer structure.** The main lobe angle is tested in the C-band wavelength (1530nm to 1565nm). The red line is for the structure with a $20\mu m$ delay length; the beam steers from -5.37° at 1531nm to -9.80° at 1557nm. The blue line is for the $40\mu m$ delay length structure; it steers the beam from 4.15° at 1529nm to -6.92° at 1557nm. The green line is for the $60\mu m$ delay length structure; it steers the beam from 12.52° at 1527nm to -5.93° at 1559nm.

Samples with the delay length of $20\mu m$, $40\mu m$, and $60\mu m$ have been tested to validate the beam steering mechanism. Fig. 8 plots the tested beam steering capability from the 4-layer devices. The angle shown in the figure is the farfield angle of the main lobe. The red, blue, and green lines are for the samples with the delay length of $20\mu m$, $40\mu m$, and $60\mu m$, respectively. The data shows that the $20\mu m$ sample can steer the beam from -5.37° at 1531nm to -9.80° at 1557nm, corresponds to **-0.170°/nm**; the $40\mu m$ sample steers the beam from 4.15° at 1529nm to -6.92° at 1557nm, corresponds to **-0.395°/nm**; the $60\mu m$ sample steers the beam from 12.52° at 1527nm to -5.93° at 1559nm, corresponds to **-0.577°/nm**. It can be concluded that the beam can be steered linearly by wavelength tuning, and the beam steering capability is proportional to the delay length, which agrees with the equation $\Delta\delta = \Delta\lambda*DL*b$. Based on this mechanism, the beam steering capability can be manipulated to any design value from 0° to 180° per nanometer wavelength. In the case of using a laser source with a high tuning step, a lower steering capability can be selected to increase the scanning resolution. In the case of a laser source with a narrower wavelength range, a higher steering capability can be set to achieve a large field of view (FOV).

### DISCUSSION

In this work, we have proposed and demonstrated a true *3-D OPA* device with a broadband high efficiency. This example presents the possibility of the multi-waveguide-layer configuration in a PIC. The existence of multiple waveguide layers perfectly addresses the two disadvantages of using the edge couplers in a traditional SOI-based OPA device. Thanks to the multiple waveguide layers, the mode match between the fiber and the on-chip waveguide is enhanced, and the emitting beams also converge in the vertical direction. We found clear evidence to show the advantages of the multi-waveguide-layer configuration with the proof-of-concept experimental results. In addition, the beam steering capability, which is enabled by the Ω-shape delay line region, is validated experimentally. In summary, the proposed *3-D OPA* device can offer high *fiber-to-chip-to-beam* efficiency, with the beam to be 2-D converged, and with the beam steering capability to be highly sensitive (manipulatable from 0° to 180° per nm wavelength) and simply operatable (only one degree of freedom in operation). This design opens a new possibility for OPA devices.

The proposed idea in Fig. 1 has been studied and experimentally proved. However, it hasn't presented the best potential of a multi-waveguide-layer PIC. Still, using another OPA design as an example, Fig. 9 illustrates an ideal multi-layer OPA implemented with electrical contacts. In this device, individual phase shifters will be applied to every single waveguide; in such a case, the emitting beam will be 2-D steerable. In addition, since the phase in every waveguide can be controlled, the input fiber is no longer required to be single mode; multimode fibers have a much larger core diameter (for example, Thorlabs GIF625 has a core diameter of $62.5\mu m$), so they can easily power approximately 50 layers of the waveguide (in the case of a $1.3\mu m$ distance between layers).

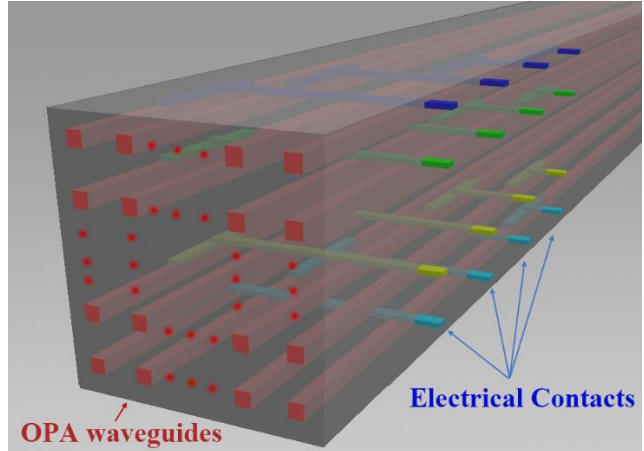

**Figure 9. Illustration of an ideal multi-layer OPA.** Waveguides are shown in red; one individual phase shifter is applied for every waveguide. The electrical contacts are colorized for different layers; they can be designed at the side wall or the top of the structure.

The fabrication strategy of such a device can also be CMOS-compatible, despite being far beyond our current fabrication capability. In the expectations, if one day such a multi-layer PIC can be readily available to researchers, a brand-new possibility will be added to the current PIC industry. The recent progress in the electronic IC industry [1-2] has presented the trend of converting from 2-D to 3-D; this development also improves the possibility of true 3-D PICs.

## MATERIALS AND METHODS
**Device Fabrication**
The samples are fabricated in the Lurie Nanofabrication Facility (LNF) in Ann Arbor, Michigan, USA. Etching multiple waveguide layers in one lithography can ensure vertical alignment between layers; it is applied in our process. In this work, samples with 1, 2, 3, and 4 layers of $Si_3N_4$ waveguide are fabricated for comparison. All four samples are fabricated with the same process introduced as follows. Firstly, all the $Si_3N_4$ waveguide and $SiO_2$ isolation layers are deposited on a blank silicon wafer. Then, the $Si_3N_4$ waveguide layers are patterned to the design shape together with the $SiO_2$ isolation layer. Next, a thick $SiO_2$ cladding layer is also deposited with PECVD. Finally, the wafer is diced into single dies, and the input and emitting sides of all dies are polished together to ensure uniformity on the surface between different samples.

It is worth noting that several device performances have been sacrificed to etch more layers within one lithography, ensuring the vertical alignment between layers. Firstly, $8\mu m$ is selected as the horizontal pitch to guarantee the etching depth. It will generate a noticeable aliasing effect in the farfield, which creates several grating lobes and limits the beam steering range of the main lobe. In addition, no samples over 4-layers can be fabricated with the current process. Since the TEOS deposition is not available, we cannot obtain a cladding layer that thoroughly covers the whole structure; for that reason, we cannot run another process to achieve the 8 waveguide layers as designed. All these sacrifices are due to the limitation of the available fabrication capability. Still, with the proof-of-concept in this work, one can easily imagine what the device can be like if it is fabricated with state-of-the-art fabrication capability.

**Measurement setup**

A 2-lens Fourier optics system is used for the measurement. The light source is tunable laser TLX1 and TLX2 from Thorlabs, a single mode fiber (SMF-28) with a fiber polarization controller is used for light coupling, and the fiber is edged coupled to the device sample. At the emitting end, an objective lens is used to monitor the nearfield of the sample. Once the nearfield pattern confirms a correct light coupling, a second lens is added to image the back focal plane (Fourier plane) to the camera, and thus the farfield pattern of the device is captured.

The coupling efficiency is measured using the testing samples. The waveguide loss is firstly measured using the cut-back method, then the sample with only two couplers and the straight connection waveguide is measured to calculate the coupling efficiency of the input coupling efficiency.

**Modeling**

In this work, commercial simulation software OmniSim (FDTD) and Lumerical (FDE & EME) are used for modeling. The Lumerical EME engine completes the design of the input coupler, and the OmniSim FDTD engine simulates the emitting end.